\documentclass[twocolumn]{aastex631}

\usepackage{soul}
\usepackage{xcolor}
\usepackage{bm}
\usepackage{amsmath}

\newcommand{\D}{\partial}

\renewcommand{\bv}[1]{\bm{#1}}
\newcommand{\vv}{\bv{v}}
\newcommand{\OO}{\bv{\Omega}}
\newcommand{\BB}{\bv{B_0}}
\newcommand{\bb}{\bv{b}}
\newcommand{\jj}{\bv{j}}
\newcommand{\OoRo}{Ro^{-1}}
\newcommand{\opara}{\hat{\bv{x}}^\Omega_\parallel}

\newcommand{\bpara}{\hat{\bv{x}}^{B_0}_\parallel}

\newcommand{\LL}{\mathcal{L}}

\def\CITE#1{{\color{red}[CITE]}}

\shorttitle{Effective drag in rotating, poorly conducting plasma turbulence}
\shortauthors{Benavides et al.}
\graphicspath{{./}}

\begin{document}

\title{Effective drag in rotating, poorly conducting plasma turbulence}

\correspondingauthor{Santiago J. Benavides}
\email{Santiago.Benavides@warwick.ac.uk}

\author[0000-0002-2281-5695]{Santiago J. Benavides}
\affiliation{Department of Earth, Atmospheric, and Planetary Sciences, 
Massachusetts Institute of Technology,
Cambridge, MA 02139, USA}
\affiliation{Mathematics Institute, University of Warwick, Coventry CV4 7AL, United Kingdom}

\author[0000-0003-4761-4766]{Keaton J. Burns}
\affiliation{Department of Mathematics, 
Massachusetts Institute of Technology,
Cambridge, MA 02139, USA}
\affiliation{Center for Computational Astrophysics, 
Flatiron Institute,
New York, NY 10010, USA}

\author[0000-0002-4366-3889]{Basile Gallet}
\affiliation{Universit\'{e} Paris-Saclay, CNRS, CEA, Service de Physique de l’Etat Condens\'{e},
91191 Gif-sur-Yvette, France}

\author[0000-0003-3589-5249]{Glenn R. Flierl}
\affiliation{Department of Earth, Atmospheric, and Planetary Sciences, 
Massachusetts Institute of Technology,
Cambridge, MA 02139, USA}

\begin{abstract} % 250 word limit

Despite the increasing sophistication of numerical models of hot Jupiter atmospheres, the large time-scale separation required in simulating the wide range in electrical conductivity between the dayside and nightside has made it difficult to run fully consistent magnetohydrodynamic (MHD) models. This has led to many studies that resort to drag parametrizations of MHD. In this study, we revisit the question of the Lorentz force as an effective drag by running a series of direct numerical simulations of a weakly rotating, poorly conducting flow in the presence of a misaligned, strong background magnetic field. We find that the drag parametrization fails once the time-scale associated with the Lorentz force becomes shorter than the dynamical time-scale in the system, beyond which the effective drag coefficient remains roughly constant, despite orders-of-magnitude variation in the Lorentz (magnetic) time-scale. We offer an improvement to the drag parametrization by considering the relevant asymptotic limit of low conductivity and strong background magnetic field, known as the quasi-static MHD approximation of the Lorentz force. This approximation removes the fast time-scale associated with magnetic diffusion, but retains a more complex version of the Lorentz force, which could be utilized in future numerical models of hot Jupiter atmospheric circulation. % 198 words

\end{abstract}


\section{Introduction} \label{sec:intro}

Hot Jupiters (HJs) are gas giant exoplanets, with masses similar to that of Jupiter, which orbit close enough to their host star that they are generally expected to be tidally locked \citep{Seager2010,HengReview2015}. The proximity to their host stars is also expected to partially ionize the upper atmospheres of HJs, leading to the interaction between the atmospheric flows and any present magnetic fields \citep{Batygin2010,Perna2010,Perna2010b,Koskinen2010,Menou2012,Koskinen2014}. Their relatively large profile when obscuring their host star, as well as their short orbital periods, make them ideal candidates for transiting observations. In the last two decades, these observations have given us access to a great amount of information about their atmospheres, including some insight into their atmospheric dynamics -- indirectly via hot spot migrations \citep{Knutson2007,Knutson2008,Zellem2014,Bell2021}
and more directly using the blue-shifting of spectra observed at the terminators \citep{Louden2015,Ehrenreich2020}. Their discovery, and the subsequent observation of atmospheric dynamics, prompted the creation of a whole sub-field devoted to the numerical modeling of the atmospheres of HJs and their close relatives. These models range in sophistication and intent, from quasi-two-dimensional shallow water models \citep{Cho2003,Langton2007,Cho2008,Showman2011,PerezBecker2013,Heng2014,Hindle2019} to three-dimensional general circulation models (GCMs) \citep{Showman2002,DobbsDixon2008,Showman2009,Menou2009,Perna2010,Rauscher2010,Heng2011,Rauscher2013,Batygin2013, Rogers2014,RogersShowman2014,Rogers2017}. 

One of the largest obstacles in modeling these atmospheres is the large conductivity contrast between the dayside and the nightside, due to the large differences in temperature \citep{Perna2010,Rogers2014,HengReview2015}. The time scale of diffusion for induced magnetic fields is proportional to this conductivity, resulting in numerical models needing to resolve small time scales in the nightside, along with large time scales to capture large-scale structures in the atmosphere. Very large time-scale separations can be impractical for numerical simulations, and, as a result, many modelers resort to a parametrization of magnetic effects that doesn't directly resolve the magnetic diffusion time-scale. The most common approach begins by assuming that any induced magnetic field is a small, rapidly diffusing perturbation around a strong background magnetic field. This leads to a time-scale associated with the Lorentz force that is equal to $\rho \sigma_e^{-1} B_0^{-2}$, where $\sigma_e$ is the electrical conductivity, $B_0$ is the strength of the background magnetic field, and $\rho$ is the fluid density \citep{Davidson1995,Davidson2013,Knaepen2008}. Modelers then substitute the Lorentz force with a drag term with an associated time-scale $\tau_{\mathrm{drag}} = \rho \sigma_e^{-1} B_0^{-2}$ \citep{Perna2010,Menou2012,Rauscher2012,Komacek2016,Koll2018,Kreidberg2018,Arcangeli2019}, which can also vary in space \citep{Rauscher2013,Beltz2021}. For small values of $\sigma_e$ the resulting drag time-scale is not restrictive for the numerics. Studies that have implemented what is dubbed as `MHD drag' have found that the structure of the atmospheric circulation significantly changes when the drag time-scale is similar to or smaller than the relevant dynamical time-scale. 

There are, however, reasons to question the validity of MHD drag. Recent attempts at modeling the full MHD equations in 3D GCMs, albeit with some simplifications, have shown that, although magnetic effects do reduce the strength of atmospheric jets (as a drag would), they also cause different morphological changes in the flow which make the authors question whether the correct prescription is a drag \citep{Rogers2014,Rogers2017Nature}. Furthermore, the authors find that the Ohmic dissipation measured in their MHD models is more than an order of magnitude smaller than would be predicted by a drag term. In an independent study, \cite{Heng2014} arrive at a similar conclusion about MHD drag by studying the shallow water MHD model. See also \cite{Potherat2017} for a similar discussion in the context of MHD experiments.
Indeed, as we'll see in the next section, when considering the approximation of the Lorentz force in the relevant limits of low conductivity and large background magnetic field, one can show that a drag-like term appears in two-dimensional flows, but in three-dimensional flows this is not the case \citep{Davidson1995,Davidson2013}. Whether this approximation can be further reduced to a drag is unclear.
The uncertainty of both the drag time-scale as well as the validity of the drag prescription itself, combined with their significant effects on the atmospheric circulation, makes this a central issue in the modeling of HJ atmospheres and impacts our understanding and expectation of their atmospheric circulation. 

In this study we \added{introduce a simplification of MHD in the low conductivity limit and }revisit the question of the Lorentz force as an effective drag. In section \ref{sec:MHD} we consider an approximation to the full MHD equations in the relevant limits of low conductivity and strong background magnetic field, called quasi-static MHD (QMHD). In particular, we focus on the form of the approximate Lorentz force and discuss its properties and potential relation to a drag-like term, whose validity we quantify using an effective drag coefficient. In section \ref{sec:methods} we describe the numerical setup and introduce an integral length-scale along the magnetic field, as a measure of the anisotropy in the flow. Then, in section \ref{sec:results} we measure the anisotropy and effective drag coefficient in a series of direct numerical simulations of QMHD turbulence in an idealized setup. We find that the drag parametrization \added{with associated time scale $\tau_{\mathrm{drag}} = \rho \sigma_e^{-1} B_0^{-2}$} works well for runs in which the dynamical time-scale is shorter than that which is associated with the Lorentz force. Beyond this, when the Lorentz force is sufficiently strong, the flow becomes anisotropic and the effective drag coefficient levels off. Finally, in section \ref{sec:conclusions} we summarize our results and propose QMHD as an intermediate model, bridging the gap between the simplicity of a drag and the complexity of the full MHD equations, to be used in future GCMs of HJs. 

\section{Rotating, weakly conducting MHD} \label{sec:MHD}

We aim to keep our fluid description of HJ atmospheres as simple as possible in an effort to focus purely on the dynamical effects of the Lorentz force. This means we will be ignoring many realistic features of HJ atmospheres, including stratification, radiation, day-night forcing contrast due to tidal locking, compressibility, and kinetic plasma effects. Given their moderate temperatures, HJ atmospheres are also likely partially ionized \citep{Batygin2010,Perna2010,Koskinen2014}. However, except at very low pressures, the ions and neutrals are expected to be highly coupled due to collisions, meaning that a single-fluid description is an appropriate characterization \citep{Perna2010,Benavides2020}. We thus begin by considering incompressible magnetohydrodynamics (MHD) with uniform density (and no buoyancy variations), subject to rotation $\OO$ and a uniform, steady background magnetic field $\BB$. \added{We are effectively considering a three-dimensional volume of a HJ atmosphere that is smaller than both the stratification scale height set by the entropy gradient and also the length scale of variation in the background magnetic field. }Further simplifications will be achieved by considering two relevant limits, low electrical conductivity, and a strong background magnetic field. 

The effect of electrical conductivity on the dynamics is quantified by the magnetic Reynolds number $Re_m$, a dimensionless parameter comparing the magnetic diffusion time-scale to the time-scale associated with the evolving flow \citep{Davidson2013}. We define $Re_m = \ell u / \eta$, where $\eta = (\mu_0 \sigma_e)^{-1}$ is the magnetic diffusivity, $\mu_0$ is the magnetic permeability, $\ell$ is a dominant length-scale of the flow, and $u$ is a velocity scale. If $Re_m \ll 1$, the diffusion and dissipation of induced magnetic fields is significant. Hot Jupiters with daysides cooler than roughly 1800 K are expected to have magnetic Reynolds number smaller than one throughout most of their atmospheres \citep{Perna2010,Hindle2021Obs}, although this assumption could break down on the dayside of the hotter HJs at lower pressures \citep{Menou2012,Rogers2014}. Importantly, dynamo instabilities are not present in flows with $Re_m \lesssim 1$, and thus do not convert kinetic energy to magnetic energy, resulting in a decaying induced magnetic field $\bb$ and a negligible Lorentz force, $\jj \times \bb$, where $\jj = \mu_0^{-1} \nabla \times \bb$, in the absence of a magnetic field generated elsewhere. 

However, in the presence of a background magnetic field (from a deeper dynamo region or from the host star), the flow can act to exchange kinetic for magnetic energy by shearing this field, inducing currents and magnetic fields. In a flow with low conductivity, the strength of this induced magnetic field scales like $b \sim Re_m B_0$, and thus the Lorentz force scales like $|\jj \times (\bb + \BB)| \sim Re_m B_0^2 \ell^{-1} \mu_0^{-1}  \propto \sigma_e B_0^2$ \citep{Davidson1995,Davidson2013,Knaepen2008}. This is the origin of $\tau_{\mathrm{drag}}$ discussed in section \ref{sec:intro}. We estimate the relevance of the Lorentz force in the dynamics by comparing the strength of the Lorentz force to the nonlinear advection term in the momentum equation, giving us our main control parameter in this study, known as the \textit{interaction parameter}:
\begin{equation}
    N \equiv Re_m \frac{B_0^2}{\mu_0 \rho u^2} = \frac{\sigma_e \ell B_0^2}{\rho u}. \label{eq:N}
\end{equation}
Despite $Re_m \ll 1$, if $B_0 / (\sqrt{\mu_0 \rho}u) $ is large enough such that $N \gtrsim 1$, then the Lorentz force can significantly affect the dynamics. Some recent studies have estimated magnetic field strengths for HJs and found magnitudes similar to that of Jupiter, but possibly up to 50 times greater for larger radius HJs, suggesting that this limit could be relevant for some HJs \citep{Reiners2010,Yadav2017,Rogers2017Nature,Hindle2021Obs}.

Flows with $Re_m \ll 1$ and $N \sim \mathcal{O}(1)$ have the distinct property that the induced magnetic field is quickly diffused away, yet the Lorentz force is \textit{not} negligible. This limit is referred to as the \textit{quasi-static} approximation to MHD (which we call `QMHD' henceforth) \citep{Moffatt1967,Sommeria1982,Davidson1995,Davidson2013,Knaepen2008}, and has been studied mainly in metallurgy and in MHD experiments due to the typically low conductivity of liquid metals \citep{Alemany1979,Sommeria1988,Gallet2009,Klein2010,Potherat2014,Baker2018}, although recent numerical studies on its turbulent properties and anisotropy have been done as well \citep{Zikanov1998,Burattini2008,Favier2010,Favier2011,Reddy2014,Verma2017}. After nondimensionalizing the equations of MHD using the uniform density $\rho$, $\ell$ and $u$, and taking the limits above, one is left with a single dynamical equation for the velocity:
\begin{eqnarray}
    \frac{\partial \vv}{\partial t} + \left(\vv \cdot \nabla \right)\vv &=& -\nabla p^* - Ro^{-1} \hat{\bv{x}}_\parallel^\Omega \times \vv \nonumber \\
    && - N \nabla^{-2}(\hat{\bv{x}}_\parallel^{B_0} \cdot \nabla)^2 \vv + \bv{F}, \label{eq:QMHD}
\end{eqnarray}
where $p^*$ is the total pressure modified by rotation and magnetic pressure, $Ro^{-1} \equiv 2\Omega \ell / u $ is the inverse Rossby number (quantifying the relative strength of the Coriolis force), $\hat{\bv{x}}_\parallel^\Omega$ and $\hat{\bv{x}}_\parallel^{B_0}$ are unit vectors in the direction of rotation and the background magnetic field, respectively, and $\bv{F}$ is a generic forcing term that can include dissipation such as viscosity and a body force (to be specified in section \ref{sec:methods}). The background magnetic field is fixed in time and is uniform in space, such that $\nabla \times \BB = B_0 \nabla \times \hat{\bv{x}}_\parallel^{B_0} = 0$. Care must be taken if considering a spatially-dependent background magnetic field $\BB(\bv{x})$, as the resulting equation will not be the same. See the discussion in section \ref{sec:conclusions}.
This equation is accompanied with the incompressibility condition $\nabla \cdot \vv = 0$. The induced magnetic field can be found using a diagnostic relation:
\begin{equation}
   \bb = -\nabla^{-2} \left(\hat{\bv{x}}_\parallel^{B_0} \cdot \nabla\right) \vv,
\end{equation}
which would be $\bb = -\nabla^{-2} \left(\BB \cdot \nabla\right) \vv / \eta$ in dimensional variables.

\begin{figure*}
	\centering{
		\includegraphics[width=0.48\textwidth]{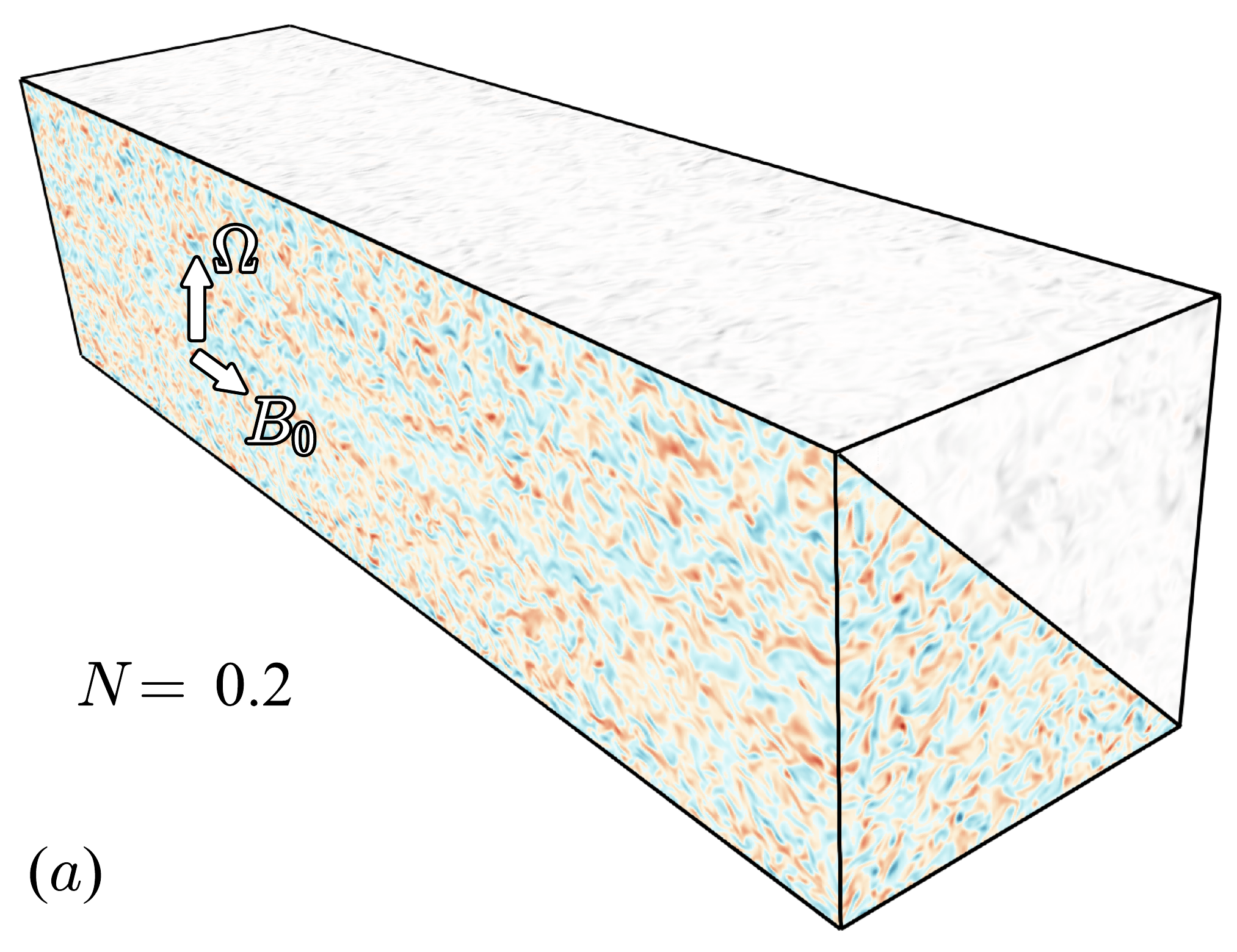}
		\includegraphics[width=0.48\textwidth]{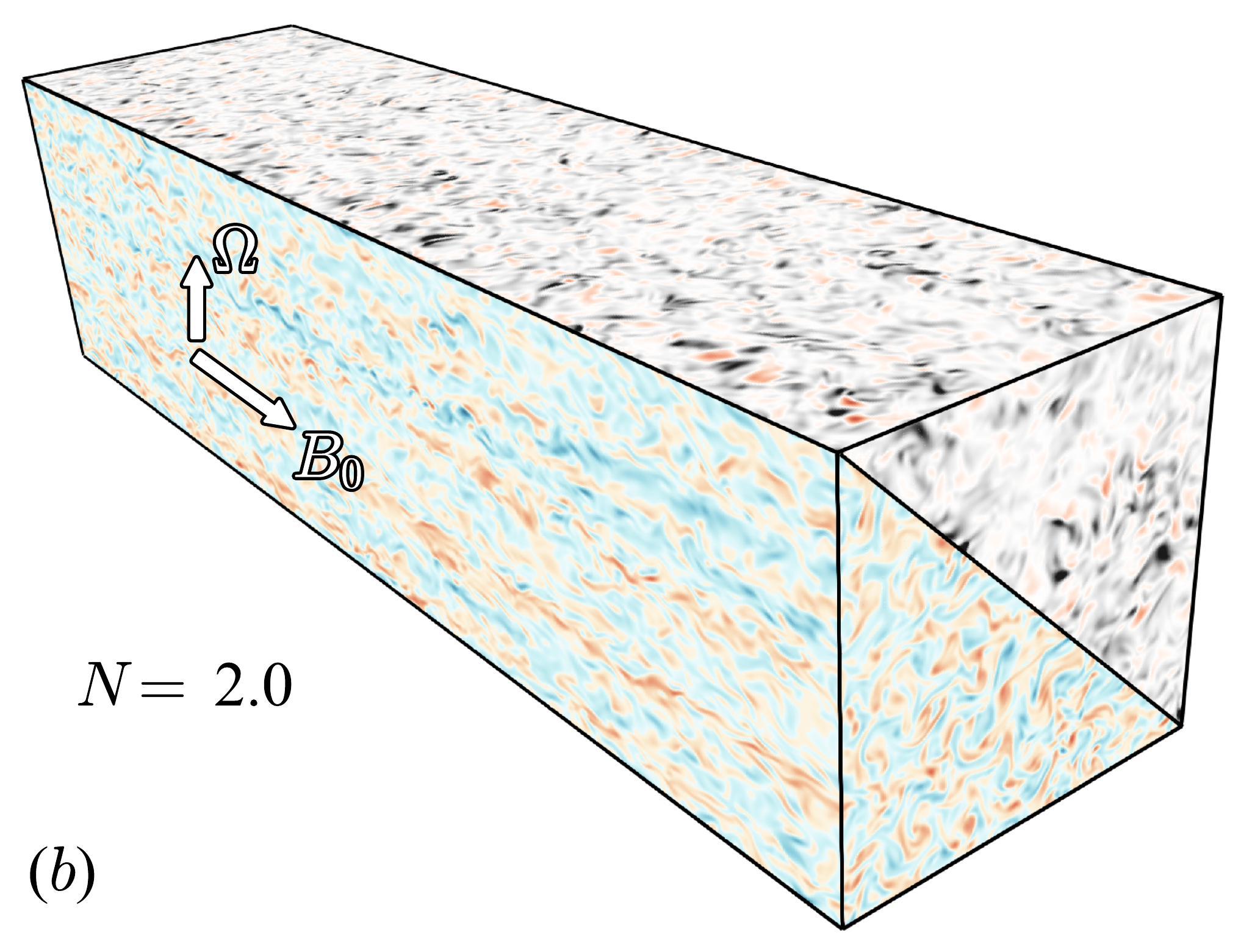}\\
		\includegraphics[width=0.48\textwidth]{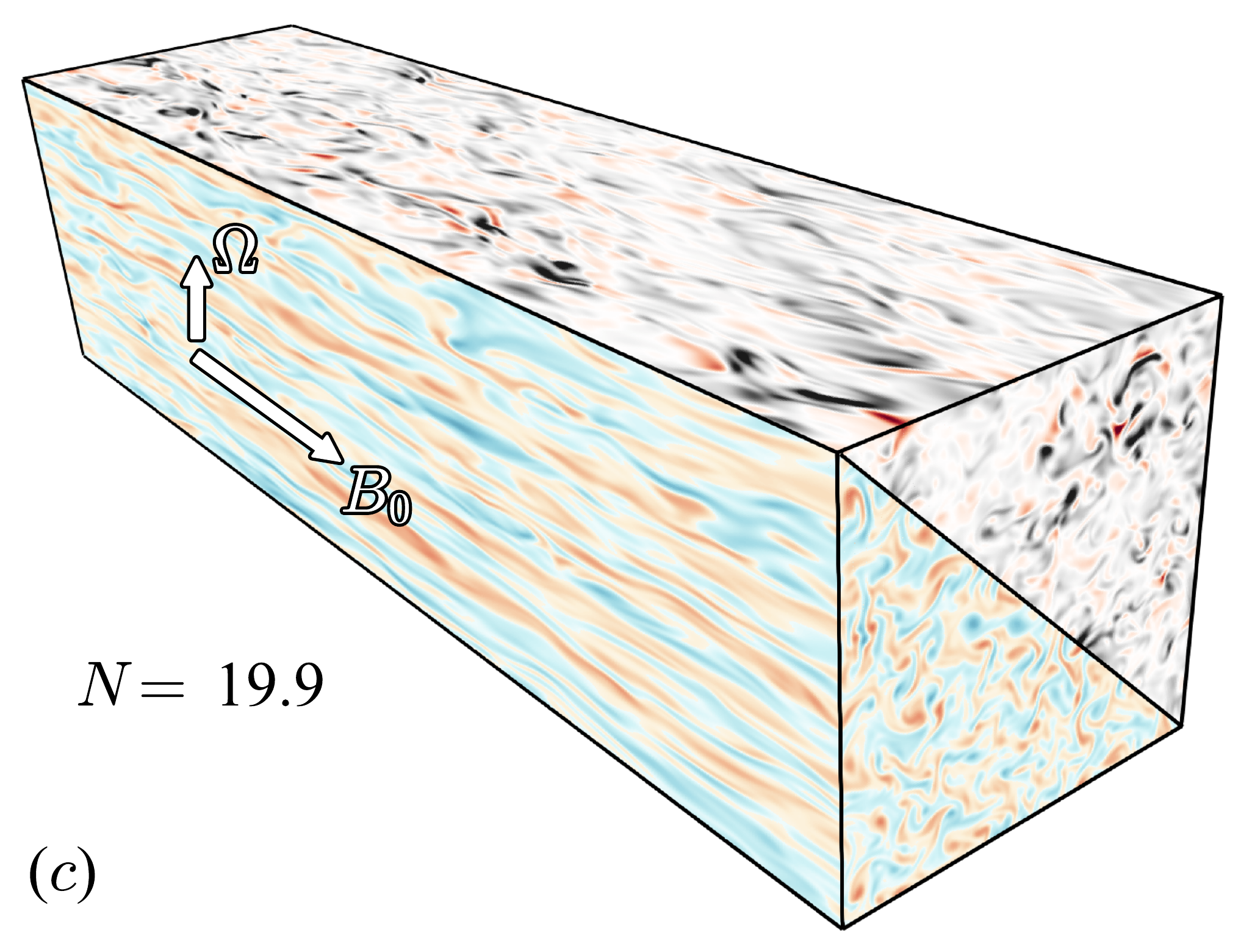}		\includegraphics[width=0.48\textwidth]{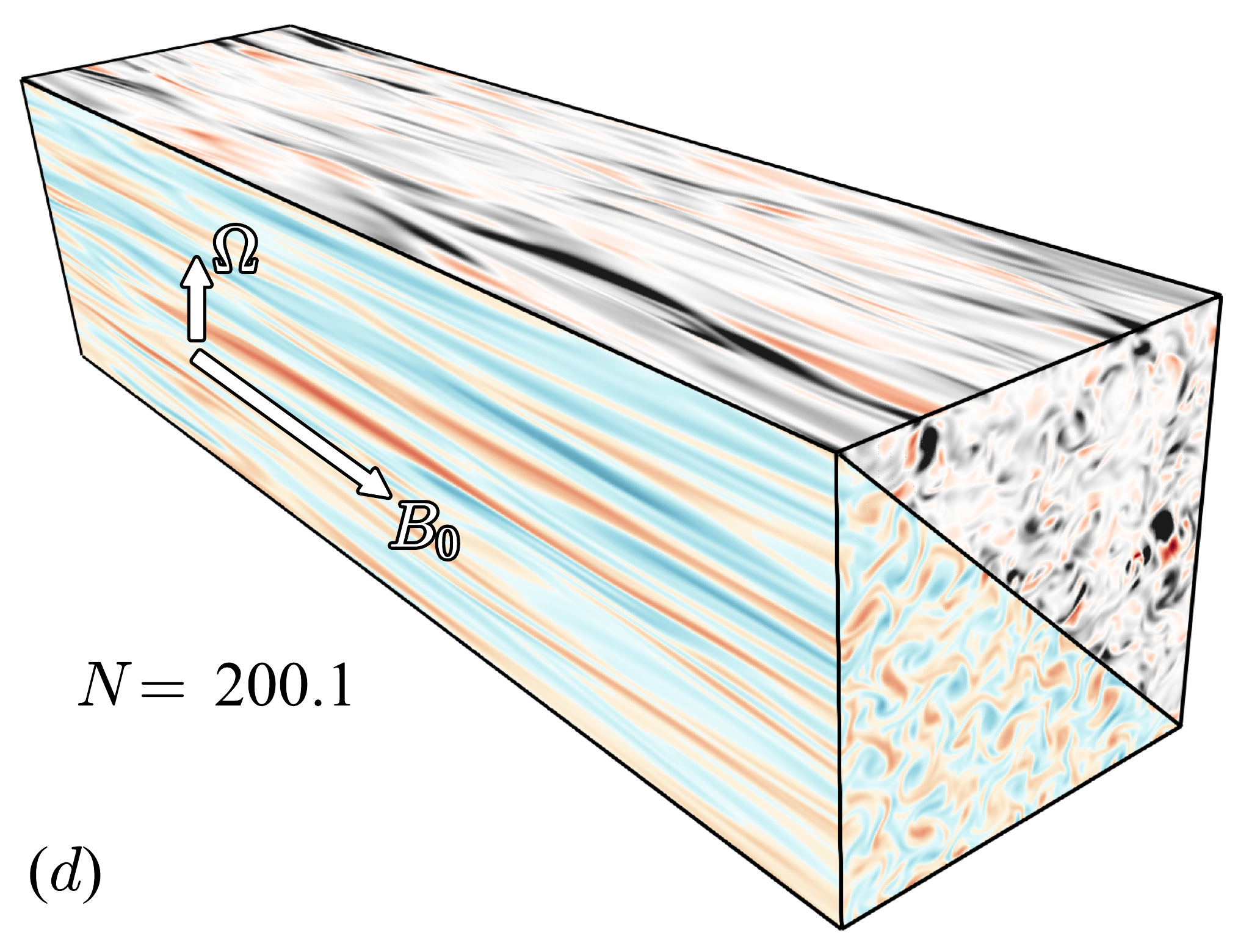}
	}
	\caption{
		Snapshots from the $L_x = 8 \pi L$ runs (Table \ref{tab:runs}) of the field-aligned vorticity $\omega = \hat{x}\cdot\left(\nabla \times \vv \right)$ (lower left, in blue and red) and the Ohmic dissipation $\vv \cdot \LL(\vv)$ (upper right, in black and red) for increasing values of the interaction parameter $N$ (equation (\ref{eq:N})). Figures \ref{fig:flow3D}(\textit{b})-(\textit{d}) represent runs with approximately equal values of volume-averaged Ohmic dissipation and $D_{\mathrm{eff}}$ (equation (\ref{eq:Deff})). The red colors represent positive values whereas the blue and black represent negative values. All snapshots use the same colorbar scale for each given field. }
	\label{fig:flow3D}
\end{figure*}
In QMHD, the Lorentz force operator, 
\begin{equation}
 \LL(\vv) \equiv - N \nabla^{-2}(\hat{\bv{x}}_\parallel^{B_0} \cdot \nabla)^2 \vv, \label{eq:LL}
\end{equation}
acts to dissipate kinetic energy from any motion which varies along the direction of the background magnetic field. In two-dimensional flows, with an in-plane $\BB$, it can be shown that $\LL(\vv)$ becomes $\LL_{2D}(\vv) = -N \tilde{\vv}_\perp^{B_0}$ (up to an irrelevant gradient field), where $\vv_\perp^{B_0}$ is the projection of the velocity perpendicular to the background field and $\tilde{\vv} \equiv \vv - L_\parallel^{-1} \int \vv \ dx_\parallel^{B_0}$ \citep{Davidson1995,Davidson2013}. These two properties would, at first glance, seem to justify the use of a drag parametrization. However, most HJs are expected to be tidally locked, resulting in order one Rossby numbers \citep{SeagerPartV2010}, which is not low enough for strong two-dimensionalization of these flows\added{ (and we are not considering the effects of strong stratification or large aspect ratio)}. In three dimensions, there are various reasons to question the use of a drag term as an approximation to the Lorentz force in QMHD. First of all, the Lorentz force $\LL(\vv)$ acts to create \textit{anisotropy} in the flow by removing energy from motions that vary along the magnetic field, but does not affect motions that are invariant along that direction. This is in sharp contrast to a drag in which any motion is affected equally and in an isotropic way. Second, a drag term removes linear momentum from the system, whereas the Lorentz force in the QMHD approximation does not (in the absence of no-slip boundary conditions). This can have profound effects on the time evolution of the system. Finally, a drag term removes energy anywhere in space where $\vv \neq 0$, whereas $\LL(\vv)$, despite removing energy in a spatially-averaged sense, may locally inject energy into the flow (Figure \ref{fig:flow3D}).

Despite these differences, $\LL(\vv)$ does remove energy from the flow, and is linear in $\vv$, so there is hope that the much simpler prescription of $\LL$ as a drag could be approximately valid in certain regimes. One can arrive at a potential justification for this by expressing $\LL(\vv)$ in Fourier space in terms of wavenumbers parallel $(k_\parallel)$ and perpendicular $(\bv{k}_\perp)$ to the background magnetic field,  resulting in $-N \sum_k (k_\perp^2 + k_\parallel^2)^{-1} k_\parallel^2 \bv{\hat{v}}(k_\parallel,\bv{k}_\perp) e^{i (\bv{k}\cdot\bv{x})}$, where $\bv{\hat{v}}$ is the Fourier transform of $\vv$. We see that, if $\max_{\bv{k}}|\bv{\hat{v}}|(k_\parallel,\bv{k}_\perp)$ happens for $k_\parallel \geq |\bv{k}_\perp|$, then we can approximate $\LL(\vv) \approx \LL_{\mathrm{drag}}(\vv) = - c_0 N \vv$, with $c_0$ being an order one constant ranging from $c_0 \approx 1/2$, when the max occurs near $k_\parallel \sim |\bv{k}_\perp|$, to $c_0 \approx 1$, when the max occurs for $k_\parallel \gg |\bv{k}_\perp|$. The latter might be relevant in a thin atmosphere where the background magnetic field projects significantly onto the thin direction and the flow contains large horizontal structures perpendicular to the background magnetic field.

In order to investigate the validity of the Lorentz force as a drag in a quantitative way, we introduce an \textit{effective drag coefficient} $D_{\mathrm{eff}}$,
\begin{equation}
    D_{\mathrm{eff}} \equiv - \frac{\langle \vv \cdot \LL(\vv) \rangle}{\langle |\vv|^2 \rangle}, \label{eq:Deff}
\end{equation}
where $\langle \cdot \rangle$ denotes a temporal and spatial average at steady state. If $\LL(\vv)$ does indeed act like a drag, at least in a volume-averaged sense, then we would expect $D_{\mathrm{eff}} \propto N$. However, if this is not the case, then $D_{\mathrm{eff}}$ will deviate from $N$. $\langle \vv \cdot \LL(\vv) \rangle$ is the Ohmic dissipation rate, so that $D_{\mathrm{eff}}$ represents the ratio of Ohmic dissipation to (twice) the kinetic energy in the flow. 

In the next section we introduce a series of direct numerical simulations of QMHD turbulence which we use to investigate how $\LL$ acts to create anisotropy, and how that, in turn, affects the validity of $\LL$ approximated as a drag. In particular, we use these simulations to determine how the effective drag coefficient depends on the interaction parameter $N$.

\section{Methods}\label{sec:methods}
We performed direct numerical simulations of the QMHD system, equation (\ref{eq:QMHD}), in a triply-periodic domain using a modified version of the Geophysical High-Order Suite for Turbulence (GHOST) \citep{Mininni2011,Benavides_Code}, a pseudo-spectral code with a fourth-order Runge–Kutta scheme for time integration and a two-thirds dealiasing rule. 
The generic forcing term $\bv{F}$ comprised of a  `hyper'-viscous term, $-\nu \nabla^4 \vv$, which acts to dissipate kinetic energy at the smallest scales, and a body forcing term $\bv{f}$, which is random (white-in-time) and injects energy into the flow at a constant rate \replaced{$f^2_0$}{$\varepsilon$} and at a single length-scale $\ell_f$, both of which are input parameters \citep{Chan2012}.
The hyper-viscosity lets us use lower resolutions while maintaining numerical accuracy, and has been shown to have no significant effect on the turbulent properties of 3D turbulence as long as the power of the gradient is small enough, as is the case here \citep{Agrawal2020}. 

The axis of rotation was chosen to be in the direction $\opara = \hat{\bv{z}}$, whereas the direction of the background magnetic field was chosen to be perpendicular to it, in the $x$-direction, $\bpara = \hat{\bv{x}}$. 
% Maybe here mention Rossby number?? S
The misalignment between the rotation axis and background field is supposed to reflect a generic case, since there is no reason to expect alignment between the two outside the dynamo region (e.g., a dipole field from the interior dynamo region). 
In some cases misaligned rotation and background field can have significant consequences on the dynamics \citep{benavides2022}, so we performed runs at other misalignment angles, $\theta$, defined to be the angle the background field makes with the rotation axis (Table \ref{tab:runs}). These runs show that our results do not depend strongly on the misalignment angle, as long as it's not zero or small.

Since $\LL$ dissipates motion that varies along $x$, we expect anisotropy to develop in our domain, manifested by structures along the $x$-direction which are larger than in the perpendicular directions. In order to accommodate the form of $\LL$, and the resulting anisotropy, we make a few specific choices in our implementation. First of all, we perform the same set of runs for various values of $L_x$, the domain size in the $x$-direction. The domain size in the perpendicular direction is fixed at $L_y = L_z = 2 \pi L$, whereas we perform sets of runs with $L_x = 2 \pi L, 4\pi L$ and $8\pi L$. 
Second, the forcing function $\bv{f}$ `stirs' the fluid in a manner that does not vary along the $x$-direction, so that the forcing does not project onto $\LL$, which would immediately dissipate the energy being forced. Since such two-dimensional forcing results in three-dimensional instabilities, the resulting flow is still approximately isotropic when $N \ll 1$.

The numerical model is nondimensionalized by $L$ and \replaced{$f_0$}{$\varepsilon$}, such that the domain size in the directions perpendicular to $\BB$ is $2 \pi$ and the forcing function has an injection rate equal to 1. In all of our runs, we have $\nu = 2 \times 10^{-6}$, $\Omega = 2$, and $\ell_f = 2 \pi / k_f$, where $k_f = 9$ (we randomly force modes $\bv{k}$ such that $8 < |\bv{k}| < 10$, making $k_f = 9$). The dominant length-scale in the problem is $\ell = \ell_f/2$, if we consider the size of a typical vortex produced by the forcing. The velocity scale is defined using \replaced{$u = (f^2_0 \ell)^{1/3} = \ell^{1/3} = (\pi / 9)^{1/3}$}{$u = (\varepsilon \ell)^{1/3} = \ell^{1/3} = (\pi / 9)^{1/3}$}. These values of $\ell$ and $u$ result in an inverse Rossby number of $\OoRo = 1.98$, and an `effective' Reynolds number of $Re = 14973$, where $Re \equiv u  \ell^3 / \nu$ is based on the hyper-dissipation used in our model.\added{ Although we do not know typical values for viscosity on hot Jupiters, given the high temperatures present, we expect it to be very small. Furthermore, with scales of motion on the order of thousands of kilometers and velocities on the order of meters per second or more, we believe that the Reynolds numbers will be much larger than one. We thus expect small scale instabilities (e.g. shear instabilities) as well as buoyancy-based instabilities to be present, leading to turbulence.} All of our runs have a resolution of 256 in each direction perpendicular to the background field. See Table \ref{tab:runs} for a list of the runs performed in our study. Note that, although we are varying $N$ over many orders of magnitude (including possibly $N \gg 1$), QMHD is the limit of MHD with vanishing magnetic Reynolds number, which always holds true. Therefore, $N\gg 1$ requires $B_0 / (\sqrt{\mu_0 \rho}u) $ to be much larger than one.
\begin{table}
\vskip12pt
\centering{
\begin{tabular}{|c|c|c|c|c|} 
    \hline
    $L_x$     & Nx  &            $\theta$            &  $N$ & \# of Runs \\ 
    \hline
    $2 \pi L$ & 256  &  $30^\circ,60^\circ,90^\circ$ & $2\times 10^{-3} - 200$ & 33 \\
    \hline
    $4 \pi L$ & 512  &            $90^\circ$         & $2\times 10^{-1} - 300$ & 8\\
    \hline
    $8 \pi L$ & 1024 &            $90^\circ$         & $2\times 10^{-1} - 600$ & 8\\
    \hline
\end{tabular}}
    \caption{A summary of the runs performed for this work. All runs have (in simulations units, nondimensionalized by $L$ and \replaced{$f_0$}{$\varepsilon$}) $\nu = 2 \times 10^{-6}$, $\Omega = 2$, and $\ell_f = 2 \pi / k_f$, where $k_f = 9$. This corresponds to an inverse Rossby number of $\OoRo = 1.98$, and an `effective' Reynolds number of $Re = 14973$. Nx represents the number of grid points in the $x$-direction. All runs have 256 grid points in both $y$ and $z$ directions.}
    \label{tab:runs}
% \vskip12pt
\end{table}

Each simulation is run until a statistical steady-state is reached, at which point the time-averages are taken until the error estimate (considering covariance) levels off. For each run we calculate the effective drag coefficient $D_{\mathrm{eff}}$, as well as an integral length-scale in the $x$-direction, defined as:
\begin{equation}
    \overline{\ell_x} = \frac{\int E(k_x) \ dk_x}{\int (k_x / 2 \pi) E(k_x) \ dk_x}, \label{eq:ell_x}
\end{equation}
where $E(k_x)$ is the time-averaged, one-dimensional energy spectrum in the $x$-direction. Note that these integrals also include contributions from the $k_x = 0$ mode. $\overline{\ell_x}$ gives an estimate of the dominant length-scale in the $x$-direction (in units of $L$), and will be used as a quantitative measure of anisotropy developing in the domain. In an isotropic system we would expect $\overline{\ell_x}  \sim \ell_f$. 

\section{Results}\label{sec:results}
\subsection{Anisotropy}
\begin{figure}
	\centering{
		\includegraphics[width=0.48\textwidth]{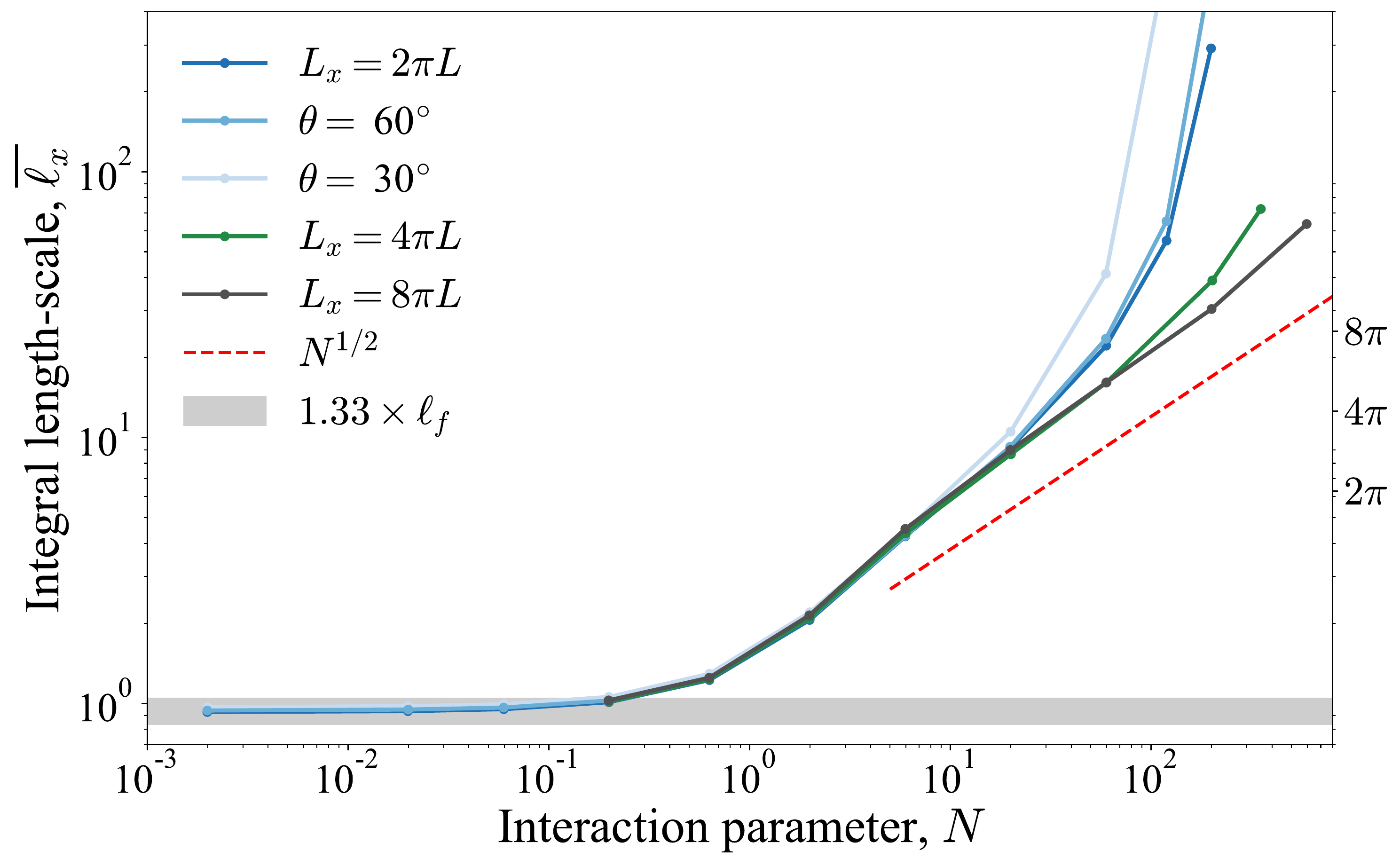}
	}
	\caption{The integral length-scale $\overline{\ell_x}$ (equation (\ref{eq:ell_x})), in units of $L$, as a function of the interaction parameter $N$ (equation (\ref{eq:N})), for various box sizes and misalignment angles. In gray we show 1.33 times the forcing length-scale range. Anisotropy develops once $N>1$, and this does not depend on the domain size or misalignment angle. For $N>1$ and sufficiently large domains which do not suffer from finite size effects (when $\overline{\ell_x} < L_x$, see axis on right), we find that $\overline{\ell_x} \sim N^{1/2}$ (red, dashed line).}
	\label{fig:ell_x}
\end{figure}

Our simulations show that significant anisotropy develops in the flow once $N\gtrsim1$ (Figures \ref{fig:flow3D} and \ref{fig:ell_x}). For $N<1$ the Lorentz force is negligible and does not affect the dynamics, resulting in approximately isotropic flow with $\overline{\ell_x} \approx 1.33 \ell_f$. The larger-than-one prefactor likely comes from the fact that the $x$-invariant structures formed by the forcing are unstable to three-dimensional perturbations at many length-scales.
Another thing to note is that the onset of anisotropy does not depend on domain size or misalignment angle.

As $N$ increases beyond one, the Lorentz force dissipates structures that vary along the $x$ direction, acting more strongly on those with small length-scales. This results in more elongated structures as $N$ increases (Figure \ref{fig:flow3D}). 
$\overline{\ell_x}$ grows until all $k_x (= k_\parallel) > 0$ modes are stable and the flow becomes exactly two-dimensional at a critical value $N_{2D}$, where $N_{2D} \propto L_x^2$ \citep{Zikanov1998,Thess2007,Favier2010,GalletDoering2015}. The effects of $N_{2D}$ are seen for $N< N_{2D}$ -- finite domain size effects appear when $\overline{\ell_x} \gtrsim L_x$ (Figure \ref{fig:ell_x}). The exact two-dimensionalization will also depend on the Reynolds number, with $N_{2D}$ increasing with $Re$ \citep{GalletDoering2015}. Given the large physical extent and Reynolds numbers of astrophysical flows, we don't expect the two-dimensionalization to be a relevant physical phenomenon.  Therefore, to avoid these effects which we believe to be irrelevant for our motivation, we consider larger domains, which allow us to push $N_{2D}$ to larger values, and therefore begin to approach the astrophysically-relevant regimes. 

The runs on larger domains reveal a power-law dependence of $\overline{\ell_x}$ with $N$, with an observed anisotropy scaling of $\overline{\ell_x} \sim N^{1/2}$ (Figure \ref{fig:ell_x}). This agrees with previous scaling predictions, such as that by \cite{Sommeria1982} who considered $\LL$ as an along-field diffusion $\LL(\vv)\approx \LL_{\mathrm{diff}}(\vv) = \kappa \partial_x^2 \vv$, with $\kappa \sim \sigma B_0^2 \ell^2_\perp / \rho$ and $\ell_\perp \sim \ell_f$ based on our forcing. In an eddy turnover time $\tau_{\mathrm{eddy}}$, motions with horizontal extent $\ell_\perp$ would diffuse vertically with a diffusion length of $\ell_x \sim \sqrt{\kappa \tau_{\mathrm{eddy}}} \propto B_0 \ell_\perp \propto N^{1/2} \ell_\perp$. This scaling can also be arrived at by considering the wavenumbers whose turbulent time-scales match that of the Lorentz force, and are therefore damped away. This gives $k \ u(k)  \sim k^{2/3} \sim N k_x^2 / k^2$, resulting in $k^{-1}_x k_\perp^{4/3} \sim \ell_x / \ell_\perp^{4/3} \sim N^{1/2}$. The slightly different scaling for $\ell_\perp$ comes from the dependence of $\tau_{\mathrm{eddy}}$ on $\ell_\perp$, based on 3D homogeneous and isotropic turbulence assumptions \citep{FrischBook}, which is not considered in the diffusivity argument.

For $1 \ll N < N_{2D}$ the anisotropy is such that the flow is almost two-dimensional (e.g., Figure \ref{fig:flow3D} (\textit{d})). Previous studies looking at turbulent energy cascades have found that inverse energy cascades (associated with two-dimensional hydrodynamics) appear before exact two-dimensionalization \citep{Smith1996,Celani2010,Alexakis2011,Deusebio2014,Sozza2015,Benavides2017,Alexakis_Review,Pouquet2019}. However, in our runs we don't see any scale coarsening in the directions perpendicular to the background field. This is due to the rotation in the $z$-direction, coupling the horizontal and out-of-plane velocities which results in a system with a forward cascade of energy \citep{benavides2022}. If rotation were to be weaker, we would expect the formation of an inverse cascade. Indeed, this seems to be occurring for the $\theta = 30^\circ$ run, where the projection of the rotation perpendicular to the background field is smaller, resulting in weaker in-plane rotation rate. The inverse cascade for this case results in larger horizontal scales $\ell_\perp$, which we believe pushes $N_{2D}$ to lower values (Figure \ref{fig:ell_x}, light blue). On the other hand, for cases with very fast rotation, the Taylor-Proudman theorem would manifest itself as flow becoming invariant along the $z$-direction, resulting in a series of shear layers varying in the third direction, $y$, and a suppressed energy cascade \citep{benavides2022}. This latter case might be more relevant for the transition regions of gas giant planets like Jupiter and Saturn, where the conditions for QMHD are also likely satisfied, but where rotation rates are significantly larger than those of HJs.

\subsection{Effective drag}
\begin{figure}
	\centering{
		\includegraphics[width=0.48\textwidth]{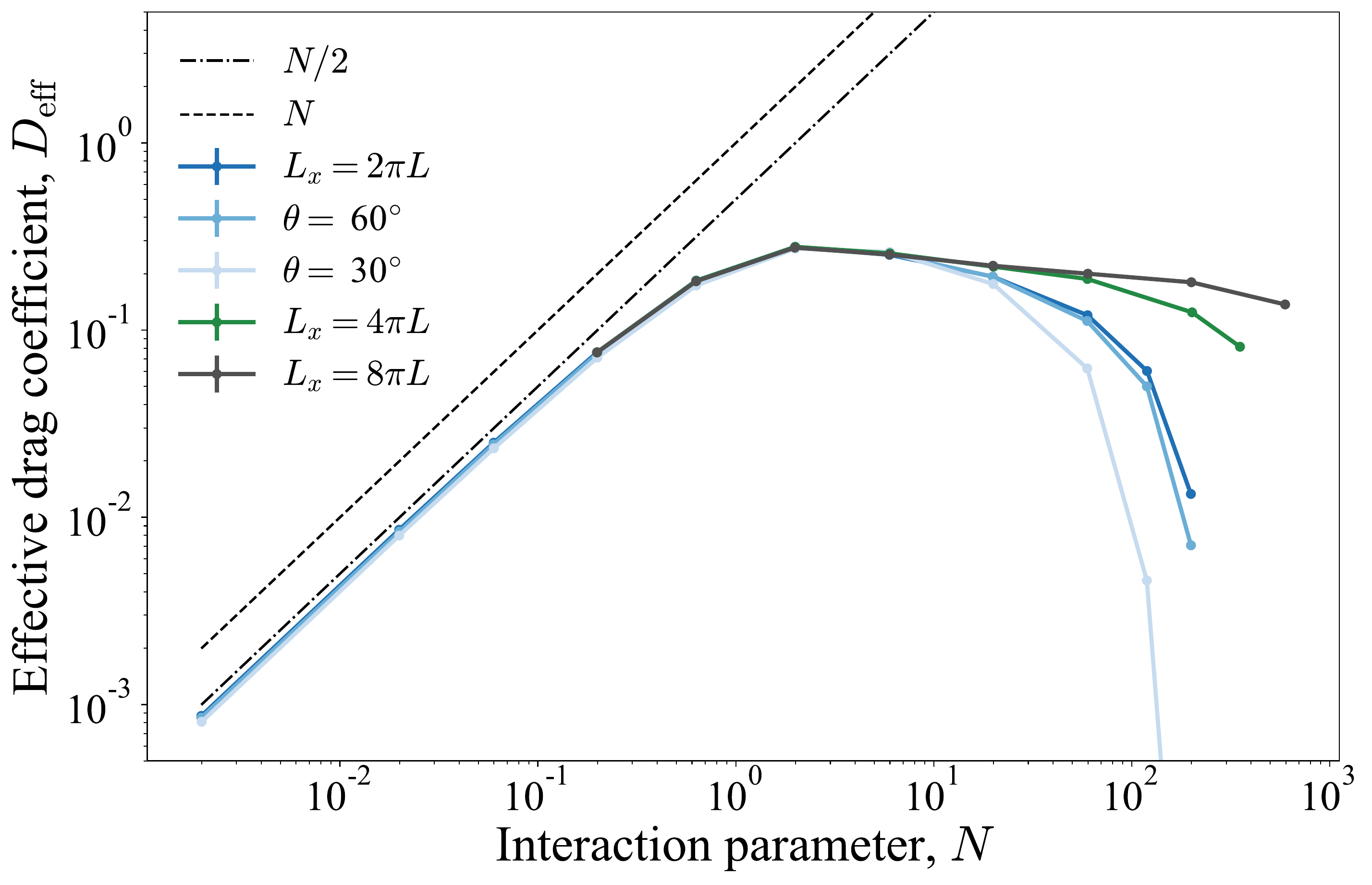}
	}
	\caption{ The effective drag coefficient $D_{\mathrm{eff}}$ (equation (\ref{eq:Deff})) versus interaction parameter $N$ (equation (\ref{eq:N})), for various box sizes and misalignment angles. For $N<1$, the effective drag coefficient is proportional to $N$ (black, dashed line), and seems to follow $N/2$ (black, dot-dashed line), suggesting that a drag formulation could be valid, with $c_0 \approx 1/2$, as expected for an energy maximum near $k_\parallel \approx |\bv{k}_\perp|$. However, the curve levels off and deviates significantly from the drag prediction by orders of magnitude when $N>1$, independent of domain size and misalignment angle $\theta$. For $N \gg 1$, the effective drag drops as the flow becomes exactly two-dimensionalized at $N_{2D}$; this limit does depend on the domain size (see section \ref{sec:results}).}
	\label{fig:eff_diss}
\end{figure}

An increase in $\overline{\ell_x}$ implies a decrease in the $x$-derivative found in $\LL$, thereby effectively lowering the Ohmic dissipation. However, the decrease in the $x$-derivative occurs as we increase $N$, which also appears in $\LL$. What is the combined effect on $D_{\mathrm{eff}}$ of $N$ increasing but the $x$-derivative decreasing? Figure \ref{fig:eff_diss} shows the effective drag coefficient $D_{\mathrm{eff}}$ as we vary the control parameter $N$.

For $N<1$, while the flow is approximately isotropic, we see a good agreement with $D_{\mathrm{eff}} \propto N$, suggesting that a drag-like parametrization could correctly capture the dynamics and Ohmic dissipation in this regime, at least in a volume-averaged sense. Indeed, $D_{\mathrm{eff}} \approx N/2$, which seems to validate the arguments made in the end of section \ref{sec:MHD} for a max energy occurring when $k_\parallel \sim |\bv{k}_\perp|$, as expected for the 3D instability of the forced 2D structures. However, as anisotropy develops for $N \gtrsim 1$, the structures that dissipate the most energy appear at larger scales (Figure \ref{fig:flow3D}) and $D_{\mathrm{eff}}$ begins to deviate from the one-to-one line. Much like the anisotropy, the deviation from the one-to-one line near $N\sim 1$ does not depend on the domain size or misalignment angle.

It is not clear \textit{a priori} what the behavior of $D_{\mathrm{eff}}$ should be beyond this point. 
For the $L_x = 2 \pi L$ runs, $D_{\mathrm{eff}}$ begins to decrease beyond $N \sim 1$. However, this is a result of the finite domain size and proximity to $N_{2D}$.
By looking at successively larger $L_x$ runs, we probe what would happen in a more realistic setting. Figure \ref{fig:eff_diss} suggests that, for large $L_x$, the effective drag coefficient $D_{\mathrm{eff}}$ levels off and remains roughly constant, despite orders of magnitude increase in $N$. 
We found that this behavior and value of $D_{\mathrm{eff}}$ does not depend strongly on $Re$. % (Figure \ref{fig:eff_diss_Re}).
Figure \ref{fig:flow3D} (\textit{b})-(\textit{d}) shows snapshots of runs which have approximately the same effective drag coefficient, while representing three orders of magnitude for $N$. Structures change in such a way so as to keep $D_{\mathrm{eff}}$ roughly constant, given the increase in $N$.

Given our findings from Figure \ref{fig:ell_x}, we can see why this behavior is a result of the anisotropy scaling $\overline{\ell_x} \sim N^{1/2}$. Combining equations (\ref{eq:LL}) and (\ref{eq:Deff}), we can re-frame $D_{\mathrm{eff}}$ as a velocity-weighted average of wavenumbers: 
\begin{equation}
    D_{\mathrm{eff}} = N \overline{\left(\frac{k^2_x}{k^2}\right)}, \label{eq:Deff2}
\end{equation}
where $k^2 = k_x^2 + k_y^2 + k_z^2$. When $N>1$, we would expect $k^2_x < k^2_\perp = k_y^2 + k_z^2$, so that we can replace $k^2$ with $k^2_\perp$ in equation (\ref{eq:Deff2}). Assuming $k_\perp$ doesn't vary significantly, since the forcing remains the same and no large-scale structures form in the flow, we can approximate it with $k_\perp \sim 2 \pi \ell_f^{-1}$. Finally, substituting $\overline{k^2_x} = (2 \pi \overline{\ell_x})^{-2}$, we end up with:
\begin{equation}
    D_{\mathrm{eff}} \approx N \left(\frac{\overline{\ell_x}}{\ell_\perp}\right)^{-2} \sim N N^{-1} \sim \text{const.}
\end{equation}
In other words, the effective drag coefficient is $N$ divided by the anisotropy of the flow squared. Since we know how the anisotropy scales with $N$, we can get $D_{\mathrm{eff}}$ as a function of $N$, which in the anisotropic limit of $1\ll N \ll N_{2D}$ turns out to be a constant.

\begin{figure}
	\centering{
		\includegraphics[width=0.48\textwidth]{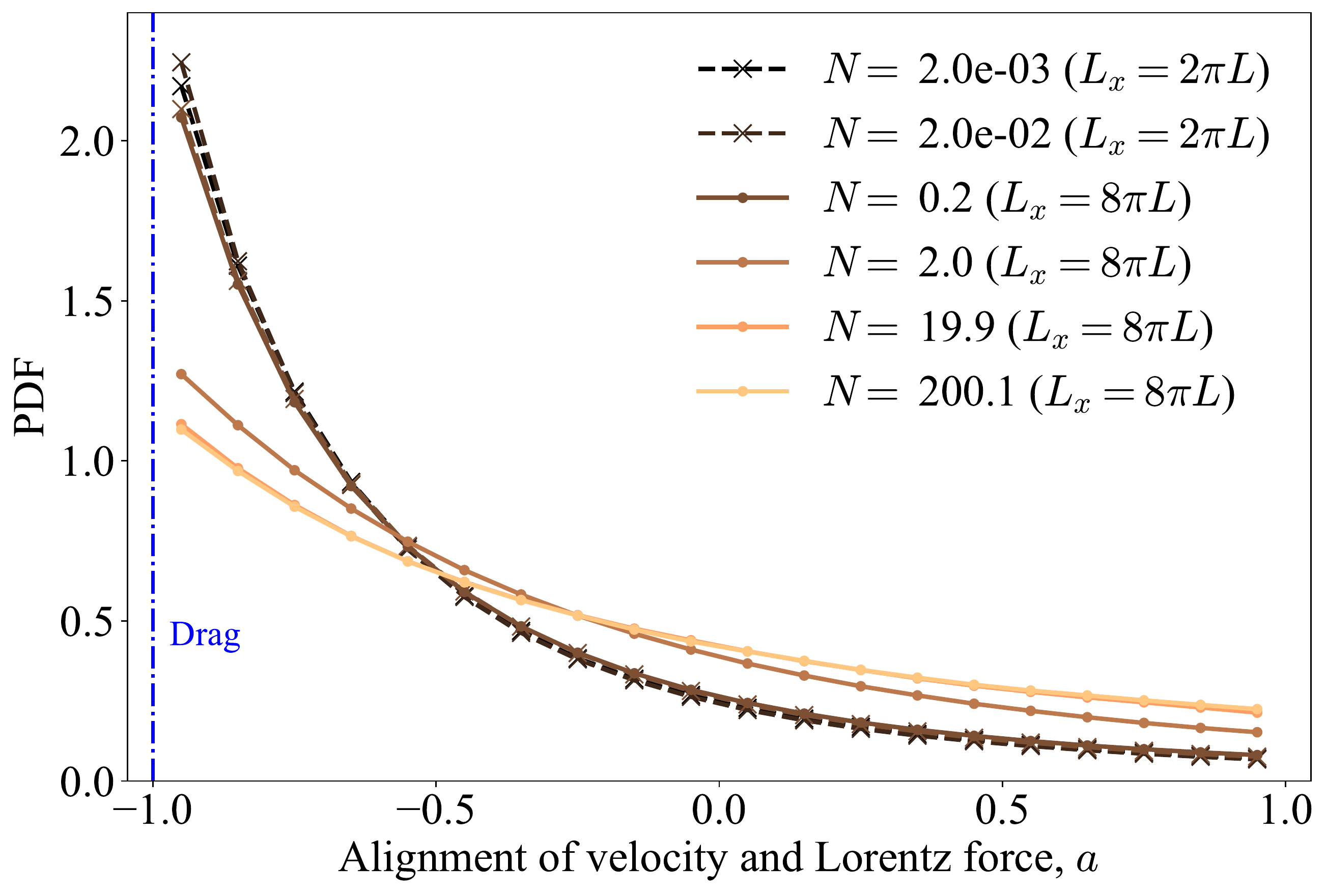}
	}
	\caption{The probability density function (PDF) in space of the alignment of velocity and the Lorentz force in terms of the dot product $a \equiv \vv \cdot \LL(\vv)/(|\vv||\LL(\vv)|)$, for the snapshots shown in Figure \ref{fig:flow3D}, as well as some runs with lower $N$. Values of $a < 0$ act to locally dissipate energy, whereas $a > 0$ act to add energy. A drag term would result in $a = -1$ at all points in space (blue, dot-dashed line). Although the effective drag coefficient would have us believe that the arguments used to justify $\LL$ as a drag are valid, this figure shows that there are still aspects of this approximation which do not hold even for $N < 1$.}
	\label{fig:align}
\end{figure}

Recall that $D_{\mathrm{eff}}$ is proportional to the Ohmic dissipation rate, so that a levelling-off of $D_{\mathrm{eff}}$ also represents the same behavior for the Ohmic dissipation\footnote{The kinetic energy is approximately the same for all runs here, partly due to the presence of a forward cascade. This might change in the case of weaker rotation and subsequent formation of an inverse cascade.}. A similar plateauing behavior has been observed for the Ohmic dissipation in HJ GCM runs with increasing temperatures \citep{Rogers2014}. The authors attribute this to a change in dynamics as $Re_m$ increases past one at lower pressures. However, they note that at higher pressures, and for some of the lower temperature runs where the plateau begins, $Re_m$ is still low, so it's possible that the QMHD effects seen here might be present and partially responsible for what is observed. 

Although Figure \ref{fig:eff_diss} seems to suggest that a drag prescription for $\LL$ is valid for $N < 1$, we emphasize that the effective drag coefficient is only a volume-averaged measure of this validity. As mentioned in section \ref{sec:MHD}, unlike a true drag term, the Lorentz force operator only dissipates the volume-averaged energy and can locally cause energy increase in certain regions, similar to a diffusion term. This fact is visible in Figure \ref{fig:flow3D} and is quantified in Figure \ref{fig:align}, which shows, for a single snapshot in time, the probability density function (PDF) in space of the alignment of velocity and the Lorentz force in terms of the dot product $a \equiv \vv \cdot \LL(\vv) / (|\vv| |\LL(\vv)|)$ for the four snapshots seen in Figure \ref{fig:flow3D}, as well as for some runs with lower $N$. A true drag term would result in a delta function distribution around $a = -1$ (blue, dot-dashed line). However, Figure \ref{fig:align} shows values of $a$ between $-1$ and $1$ even for the runs with $N<1$, suggesting both positive and negative contributions to the local energy balance due to the Lorentz force, while clearly showing overall energy dissipation given by the peaking PDFs for $a=-1$. All of this suggests that, although a drag might capture the energy balance correctly\added{ (albeit, only for $N<1$)}, this approximation will not result in the same spatial structures -- a drag is not the same as an along-field diffusion. That said, large-scale HJ atmosphere simulations might not resolve along-field scales very well. If one considers under-resolving as an effective averaging of such scales, then our results could justify the implementation of an effective drag.

\section{Conclusions}\label{sec:conclusions}
In this study, we considered the turbulent dynamics of rotating MHD in the presence of a background magnetic field, in the combined limit of $Re_m \ll 1$ and $B_0/(u\sqrt{\mu_0 \rho}) \gg 1$, termed quasi-static MHD (QMHD). Motivated by approaches used in the study of hot Jupiter (HJ) atmospheres, we have shown that a drag parametrization of the Lorentz force operator $\LL$ fails once the ratio of the dynamical timescale to the Lorentz timescale, quantified by the interaction parameter $N$, is larger than one. This happens because the Lorentz force dissipates structures that vary along the background field, creating anisotropy in the flow, which in turn acts to reduce the Ohmic dissipation, thereby reducing the effective drag. The development of anisotropy with increasing $N$ is such that the effective drag coefficient remains constant for $N>1$, despite $N$ varying by orders of magnitude. The levelling off of $D_{\mathrm{eff}}$ for $N>1$ has significant implications for simulations parametrizing $\LL$ as a drag, since we see values of $D_{\mathrm{eff}}$ deviating by orders of magnitude from what would be predicted if one assumes $\LL(\vv) = \LL_{\mathrm{drag}}(\vv) = -c_0 N \vv$. This could also result in severely overestimating the amount of Ohmic dissipation, as well as misrepresenting the true dynamics of HJ atmospheres. Although we have shown more generally that the drag prescription does not fully represent the form of the Lorentz force at low conductivities, \added{we are aware of the various difficulties in performing a GCM run with full MHD, motivating the use of MHD drag. It would be of interest to see if our results on effective drag carry through to full MHD or QMHD GCMs, and if the implementation of an MHD drag of the form $D_{\mathrm{eff}}(\bv{x}) = \min\{N(\bv{x}), N_0\}$, with $N_0$ being a constant of order one, is valid.}

The main motivation for a drag parametrization of the Lorentz force comes from the restrictively small time scales associated with very large magnetic diffusivities (low electrical conductivity). The drag time-scale is much larger than the time-scale associated with magnetic diffusion, allowing modellers to bypass this problem. While we have found that a drag parametrization fails for $N>1$, we want to emphasize that the same time-scale advantage exists in the QMHD limit, despite the more complicated form of the operator associated with the Lorentz force. This will hopefully motivate the use of the QMHD approximation in models of HJ atmospheres. Even for the case of a spatially-dependent background magnetic field or conductivity, its implementation would be straight forward if one considers separately the Lorentz force $\mu_0^{-1} (\nabla \times \bb)\times \BB(\bv{x})$ and the induced magnetic field $\bb = - \nabla^{-2} (\eta(\bv{x})^{-1} \nabla \times (\vv \times \BB(\bv{x})))$. An alternative which might be easier to implement would be to approximate $-\nabla^{-2}$ with some horizontal length-scale $\ell_\perp^2$, similar to what was done when considering $\LL(\vv)$ as an along-field diffusivity, $\LL(\vv)\approx \LL_{\mathrm{diff}}(\vv) = \kappa \partial_x^2 \vv$, with $\kappa \sim \sigma B_0^2 \ell^2_\perp / \rho$ \citep{Sommeria1982}. Although the expression for $\LL$ (equation (\ref{eq:LL})) would be modified in the presence of a spatially-dependent background magnetic field, we expect our results to hold for those cases, as well. 

We thank Thaddeus D. Komacek for insightful discussions and helpful suggestions. This research was carried out in part during the 2019 Summer School at the Center for Computational Astrophysics, Flatiron Institute. The Flatiron Institute is supported by the Simons Foundation.
SJB acknowledges funding from the National Aeronautics and Space Administration (Award Number: 80NSSC20K1367) issued through the Future Investigators in NASA Earth and Space Science and Technology (NNH19ZDA001N-FINESST) within the NASA Research Announcement (NRA): Research Opportunities in Space and Earth Sciences (ROSES-2019).

\software{Geophysical High-Order Suite for Turbulence (GHOST). Branch: pre-release-2, Commit: 979b9fee3425dd7836687ffe6acff66e41d7083f \citep{Benavides_Code}. Forked and modified from \cite{Mininni2011}, see \url{https://github.com/pmininni/GHOST}.
          }

\bibliography{main}{}
\bibliographystyle{aasjournal}

\listofchanges

\end{document}